\documentclass[prd,showpacs,showkeys,nofootinbib,twocolumn]{revtex4-1}

\usepackage{amsmath,amsfonts,amssymb}
\usepackage{graphicx}

\begin{document}

\title{Conservation laws in gravity: A unified framework}

\author{Yuri N. Obukhov}
\email{obukhov@ibrae.ac.ru}
\affiliation{Theoretical Physics Laboratory, Nuclear Safety Institute, 
Russian Academy of Sciences, B.Tulskaya 52, 115191 Moscow, Russia} 

\author{Dirk Puetzfeld}
\email{dirk.puetzfeld@zarm.uni-bremen.de}
\homepage{http://puetzfeld.org}
\affiliation{ZARM, University of Bremen, Am Fallturm, 28359 Bremen, Germany} 

\date{ \today}

\begin{abstract}
We study general metric-affine theories of gravity in which the metric and connection are the two independent fundamental variables. In this framework, we use Lagrange-Noether methods to derive the identities and the conservation laws that correspond to the invariance of the action under general coordinate transformations. The results obtained are applied to generalized models with nonminimal coupling of matter and gravity, with a coupling function that depends arbitrarily on the covariant gravitational field variables. 
\end{abstract}

\pacs{04.50.-h; 04.20.Fy; 04.20.Cv}
\keywords{Metric-affine gravity, Conservation laws; Noether theorem; Variational methods}


\maketitle

\section{Introduction}\label{introduction_sec}

A metric and connection are the two fundamental geometrical objects on a spacetime manifold. They play an important role in the description of gravitational phenomena in the framework of what can be quite generally called an Einsteinian approach to gravity. The principles of equivalence and general coordinate covariance are the cornerstones of this approach. As Einstein himself formulated, the crucial achievement of his theory was the elimination of the notion of inertial systems as preferred ones among all possible coordinate systems. 

In Einstein's general relativity (GR) theory, gravitation is associated with the metric tensor alone. Nevertheless, it is worthwhile to stress that Einstein clearly understood the different physical statuses of the metric and the connection: ``at first Riemannian metric was considered the fundamental concept on which the general theory of relativity and thus the avoidance of the inertial system were based. Later, however, Levi-Civita rightly pointed out that the element of the theory that makes it possible to avoid the inertial system is rather the infinitesimal displacement field $\Gamma_{ik}{}^j$. The metric or the symmetric tensor field $g_{ik}$ which defines it is only indirectly connected with the avoidance of the inertial system insofar as it determines a displacement field.'' (Appendix II in \cite{Einstein:1956}).

There exists a variety of gravitational theories that generalize or extend the physical and mathematical structure of GR. Among these theories there are large classes of so-called $f(R)$ models, and of theories with nonminimal coupling to matter; they are developed in particular in the context of relativistic cosmology (but not only there), see \cite{Schmidt:2007,Straumann:2008,Nojiri:2011}. The so-called Palatini approach represents another class of widely discussed theories in which the metric and the connection are treated as independent variables in the action principle \cite{Hehl:1978,Hehl:1981,Sotiriou:2010}. Last but not least, we should mention the vast family of the gauge gravity theories constructed using a Yang-Mills type of approach \cite{Blagojevic:2002,Hehl:2013}. The formalism of metric-affine gravity makes it possible to study all these different alternative theories in a unified framework. The corresponding spacetime landscape \cite{Schouten:1954} includes as special cases the geometries of Riemann, Riemann-Cartan, Weyl, Weitzenb\"ock, etc. 

In this paper, we develop a general Lagrange-Noether framework for metric-affine gravity theories and derive the Noether identities, which correspond to the general coordinate invariance of the action. These identities are then used to derive the conservation laws. The results obtained generalize a number of findings available in the earlier literature (correcting some shortcomings) and they represent a unified framework, which is, e.g., applicable to the analysis of the equations of motion in a wide class of gravity theories, with minimal as well as with nonminimal coupling of matter to the gravitational field. Ultimately, our aim is to set up a complete scheme that is suitable for the systematic study of experimental tests of the gravitational theories. In this connection, it is worthwhile to cite Einstein \cite{Einstein:1921} again who underlined that ``[...] the question whether this continuum has a Euclidean, Riemannian, or any other structure is a question of physics proper which must be answered by experience, and not a question of a convention to be chosen on grounds of mere expediency.''

The structure of the paper is as follows: In section \ref{sec_models}, we give a short overview of the different geometrical objects and notions which are defined on the spacetime manifold and are used for the description of the gravitational field in the wide class of models under consideration. We then develop, in section \ref{Noether_sec}, a general Lagrange-Noether analysis of an arbitrary system of interacting matter and gravitational fields, invariant under general coordinate transformations. The results obtained are subsequently applied in Sec.\ \ref{conservation_sec} to a model with nonminimal coupling of gravity and matter, for which we explicitly work out the corresponding conservation laws. To demonstrate how the present formalism fits into the gauge-theoretic scheme based on the general affine group, we discuss the dynamics of the gravitational field in Sec.~\ref{MAG}. Einstein's general relativity arises as a special case in our general framework, and in Sec.\ \ref{quadrupole} we analyze nonminimally coupled matter with intrinsic moments in GR. Finally, we draw our conclusions in section \ref{conclusion_sec}.

Our notations and conventions are those of \cite{Hehl:1995}. In particular, the basic geometrical quantities such as the curvature, torsion, and nonmetricity are defined as in \cite{Hehl:1995}, and we use the Latin alphabet to label the spacetime coordinate indices. Furthermore, the metric has the signature $(+,-,-,-)$. It should be noted that our definition of the metrical energy-momentum tensor is different from the definition used in \cite{Bertolami:etal:2007,Nojiri:2011,Puetzfeld:Obukhov:2013}. 

\section{Metric-affine geometry: an overview} \label{sec_models}

\begin{figure}
\begin{center}
\includegraphics[height=6cm]{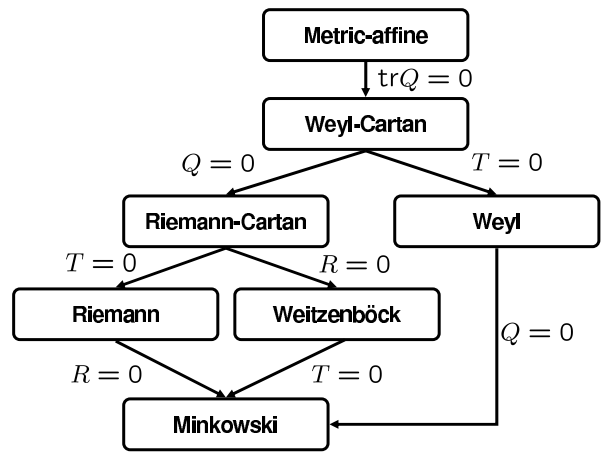}
\end{center}
\caption{Different spacetime types as special cases of a general metric-affine spacetime ($R_{kli}{}^j \neq 0$, $T_{kl}{}^i \neq 0$, $Q_{kij} \neq 0$). The abbreviations $R$ (curvature), $T$ (torsion), and $Q$ (nonmetricity) over the arrows denote the vanishing of the corresponding geometrical object.}
\label{fig:1}
\end{figure}

In the metric-affine theory of gravity (MAG), the gravitational physics is encoded in two fields: the metric tensor $g_{ij}$ and an independent linear connection $\Gamma_{ki}{}^j$. The latter is not necessarily symmetric and compatible with the metric. From the geometrical point of view, the metric introduces lengths and angles of vectors, and thereby determines the distances (intervals) between points on the spacetime manifold. The connection introduces the notion of parallel transport and defines the covariant differentiation of tensor fields. The zoo of geometric structures that arises on the spacetime is described by three fundamental objects -- the curvature, the torsion, and the nonmetricity \cite{Schouten:1954}:
\begin{eqnarray}
R_{kli}{}^j &=& \partial_k\Gamma_{li}{}^j - \partial_l\Gamma_{ki}{}^j + \Gamma_{kn}{}^j \Gamma_{li}{}^n - \Gamma_{ln}{}^j\Gamma_{ki}{}^n,\label{curv}\\
T_{kl}{}^i &=& \Gamma_{kl}{}^i - \Gamma_{lk}{}^i,\label{tors}\\ \label{nonmet}
Q_{kij} &=& -\,\nabla_kg_{ij} = - \partial_kg_{ij} + \Gamma_{ki}{}^lg_{lj} + \Gamma_{kj}{}^lg_{il}.
\end{eqnarray}
A general metric-affine spacetime ($R_{kli}{}^j \neq 0$, $T_{kl}{}^i \neq 0$, $Q_{kij} \neq 0$) incorporates several other spacetimes a special cases, see figure \ref{fig:1} for an overview.

In order to describe the deviation of the geometry from the Riemannian one, it is convenient to introduce the {\it distortion} tensor 
\begin{equation}
N_{kj}{}^i = \widetilde{\Gamma}_{kj}{}^i - \Gamma_{kj}{}^i.\label{dist}
\end{equation}
This object measures a deviation of the connection from the Christoffel symbols
\begin{equation}
\widetilde{\Gamma}_{kj}{}^i = {\frac 12}g^{il}(\partial_jg_{kl} + \partial_kg_{lj} - \partial_lg_{kj}).\label{Chr}
\end{equation}

Our aim is to study gravity models in which the Lagrangian is allowed to depend on all the geometrical variables and on the matter fields, which we collectively denote by $\psi^A$. The functional form of the total Lagrangian of the coupled system of material and gravitational fields is then given by
\begin{eqnarray}
L =  L(g_{ij}, R_{kli}{}^j, N_{kj}{}^i, \psi^A, \nabla_i\psi^A), \label{ansatz_lagrangian_model_3}
\end{eqnarray}
which may depend arbitrarily on its arguments. Obviously, $L$ can be understood as a function of independent scalar invariants constructed in all possible ways from the components of the curvature, torsion, and nonmetricity, since the latter two objects can be expressed in terms of the distortion tensor:
\begin{eqnarray}
T_{kl}{}^i &=& -\,2N_{[kl]}{}^i,\label{torsN}\\ 
Q_{kij} &=& -\,2N_{k(i}{}^lg_{j)l}.\label{nonmetN}
\end{eqnarray}

\section{General Lagrange-Noether analysis}\label{Noether_sec}

As a first step, we notice that the gravitational (geometrical) and material variables can be described together by means of a multiplet, which we denote by $\Phi^J = (g_{ij},\Gamma_{ki}{}^j,\psi^A)$. We do not specify the range of the multi-index $J$ at this stage. The matter fields may include, besides the true material variables, also auxiliary fields such as Lagrange multipliers. With the help of the latter we can impose various constraints on the geometry of the spacetime. Furthermore, we can use the Lagrange multipliers to describe models in which the Lagrangian depends on arbitrary-order covariant derivatives of the curvature, torsion, and nonmetricity. Then the general action reads 
\begin{equation}
I = \int\,d^4x\,{\cal L},\label{action}
\end{equation}
where the Lagrangian density ${\cal L} = {\cal L}(\Phi^J,\partial_i\Phi^J)$ depends arbitrarily on the set of fields $\Phi^J$ and their first derivatives. 

Our aim is to derive Noether identities that correspond to general coordinate transformations. However, it is more convenient to start with arbitrary infinitesimal transformations of the spacetime coordinates and the matter fields. They are given as follows:
\begin{eqnarray}
x^i &\longrightarrow& x'^i (x) = x^i + \delta x^i,\label{dx}\\
\Phi^J(x) &\longrightarrow& \Phi'^J(x') = \Phi^J(x) + \delta\Phi^J(x).
\label{dP}
\end{eqnarray} 
Within the present context it is not important whether this is a symmetry transformation under the action of any specific group. The total variation (\ref{dP}) is a result of the change of the form of the functions and of the change induced by the transformation of the spacetime coordinates (\ref{dx}). To distinguish the two pieces in the field transformation, it is convenient to introduce the {\it substantial variation}: 
\begin{equation}\label{sub}
\overline{\delta}\Phi^J := \Phi'^J(x) - \Phi^J(x) = \delta\Phi^J - \delta x^k\partial_k\Phi^J.
\end{equation}
By definition, the substantial variation commutes with the partial derivative, $\overline{\delta}\partial_i = \partial_i\overline{\delta}$. 

We need the total variation of the action:
\begin{equation}
\delta I = \int \left[d^4x\,\delta{\cal L} + \delta(d^4x)\,{\cal L}\right].
\end{equation}
A standard derivation shows that under the action of the transformation (\ref{dx})-(\ref{dP}) the total variation reads
\begin{equation}
\delta I = \int d^4x \left[ {\frac {\delta {\cal L}}{\delta\Phi^J}} \,\overline{\delta}\Phi^J + \partial_i\left({\cal L}\,\delta x^i + {\frac {\partial {\cal L}}{\partial(\partial_i\Phi^J)}}\,\overline{\delta} \Phi^J \right)\right]. \label{master}
\end{equation}
Here the variational derivative is defined, as usual, by
\begin{equation}\label{var}
{\frac {\delta {\cal L}}{\delta\Phi^J}} := {\frac {\partial {\cal L}}{\partial\Phi^J}} - \partial_i\left({\frac {\partial {\cal L}}{\partial(\partial_i\Phi^J)}}\right).
\end{equation}

\subsection{General coordinate invariance}

Now we specialize to the general coordinate transformations. For infinitesimal changes of the spacetime coordinates and (matter and gravity) fields (\ref{dx}) and (\ref{dP}) we have $x^i\rightarrow x^i + \delta x^i$, $g_{ij}\rightarrow g_{ij} + \delta g_{ij}$, $\Gamma_{ki}{}^j\rightarrow \Gamma_{ki}{}^j + \delta\Gamma_{ki}{}^j$, and $\psi^A \rightarrow \psi^A + \delta\psi^A$, with
\begin{eqnarray}
\delta x^i &=& \xi^i(x),\label{dex}\\ \label{dgij}
\delta g_{ij} &=& -\,(\partial_i\xi^k)\,g_{kj} - (\partial_j\xi^k)\,g_{ik},\\
\delta\psi^A&=&-\,(\partial_i\xi^j)\,(\sigma^A{}_B)_j{}^i\,\psi^B, \label{dpsiA}\\
\delta\Gamma_{ki}{}^j &=&  -\,(\partial_k\xi^l)\,\Gamma_{li}{}^j - (\partial_i\xi^l)\,\Gamma_{kl}{}^j \nonumber\\ 
&& + \,(\partial_l\xi^j)\,\Gamma_{ki}{}^l - \partial^2_{ki}\xi^j.\label{dG}
\end{eqnarray}
The four arbitrary functions $\xi^i(x)$ parametrize an arbitrary local diffeomorphism. Here, $(\sigma^A{}_B)_j{}^i$ are the generators of general coordinate transformations that satisfy the commutation relations
\begin{eqnarray}
(\sigma^A{}_C)_j{}^i(\sigma^C{}_B)_l{}^k &-& (\sigma^A{}_C)_l{}^k (\sigma^C{}_B)_j{}^i\nonumber\\ 
&=& (\sigma^A{}_B)_l{}^i\,\delta^k_j - (\sigma^A{}_B)_j{}^k \,\delta^i_l.\label{comms}
\end{eqnarray}
Substituting (\ref{dex})-(\ref{dG}) into (\ref{master}), and making use of the substantial derivative definition (\ref{sub}), we find
\begin{eqnarray}
\delta I &=& -\,\int d^4x \biggl[\xi^k\,\Omega_k + (\partial_i\xi^k)\,\Omega_k{}^i\nonumber\\
&& +\,(\partial^2_{ij}\xi^k)\,\Omega_k{}^{ij} + (\partial^3_{ijn}\xi^k)\,\Omega_k{}^{ijn}\biggr],
\label{masterG} 
\end{eqnarray}
where explicitly
\begin{eqnarray}
\Omega_k &=& {\frac {\delta {\cal L}}{\delta g_{ij}}}\,\partial_kg_{ij} + {\frac {\delta {\cal L}}{\delta\psi^A}}\,\partial_k\psi^A + {\frac {\partial {\cal L}}{\partial \Gamma_{ln}{}^m}}\partial_k\Gamma_{ln}{}^m\nonumber\\
&& + \,\partial_i\left({\frac {\partial {\cal L}}{\partial \partial_ig_{mn}}}\partial_kg_{mn}\right)+ {\frac {\partial {\cal L}}{\partial \partial_i\Gamma_{ln}{}^m}}\partial_k\partial_i\Gamma_{ln}{}^m \nonumber\\
&&+ \,\partial_i\left({\frac {\partial {\cal L}}{\partial\partial_i\psi^A}} \,\partial_k\psi^A - \delta^i_k{\cal L} \right),\label{Om1}\\
\Omega_k{}^i &=& 2{\frac {\delta {\cal L}}{\delta g_{ij}}}\,g_{kj} + {\frac {\delta {\cal L}}{\delta\psi^A}}\,(\sigma^A{}_B)_k{}^i\,\psi^B + {\frac {\partial {\cal L}}{\partial\partial_i\psi^A}}\partial_k\psi^A 
\nonumber\\ 
&& - \delta^i_k{\cal L} + \,2\partial_n\left({\frac {\partial {\cal L}}{\partial \partial_ng_{ij}}}g_{jk}\right) + {\frac {\partial {\cal L}}{\partial \partial_ig_{mn}}}\partial_kg_{mn} \nonumber\\
&& + \partial_j\!\left(\!{\frac {\partial {\cal L}}{\partial\partial_j\psi^A}}(\sigma^A{}_B)_k{}^i\psi^B\!\right)\! + \,{\frac {\partial {\cal L}}{\partial \Gamma_{li}{}^j}}\,\Gamma_{lk}{}^j \nonumber\\
&&+ {\frac {\partial {\cal L}}{\partial \Gamma_{il}{}^j}}\,\Gamma_{kl}{}^j - {\frac {\partial {\cal L}}{\partial \Gamma_{lj}{}^k}}\,\Gamma_{lj}{}^i + \,{\frac {\partial {\cal L}}{\partial \partial_i\Gamma_{ln}{}^m}}\,\partial_k \Gamma_{ln}{}^m\nonumber\\
&&  + {\frac {\partial {\cal L}}{\partial \partial_n\Gamma_{il}{}^m}} \,\partial_n\Gamma_{kl}{}^m+ \,{\frac {\partial {\cal L}}{\partial \partial_n\Gamma_{li}{}^m}}\,\partial_n \Gamma_{lk}{}^m \nonumber\\
&&  - {\frac {\partial {\cal L}}{\partial \partial_n\Gamma_{lm}{}^k}}
\,\partial_n\Gamma_{lm}{}^i ,\label{Om2}\\
\Omega_k{}^{ij} &=& {\frac {\partial {\cal L}}{\partial\partial_{(i}\psi^A}} (\sigma^A{}_B)_k{}^{j)}\psi^B + {\frac {\partial {\cal L}}{\partial \Gamma_{(ij)}{}^k}}
+ {\frac {\partial {\cal L}}{\partial \partial_{(i}\Gamma_{j)l}{}^m}}\Gamma_{kl}{}^m \nonumber\\
&&+ \,2{\frac {\partial {\cal L}}{\partial \partial_{(i}g_{j)n}}}g_{kn} + {\frac {\partial {\cal L}}{\partial \partial_{(i}\Gamma_{|l|j)}{}^m}}\,\Gamma_{lk}{}^m \nonumber\\
&& - {\frac {\partial {\cal L}}{\partial \partial_{(i}\Gamma_{|ln|}{}^k}}\,\Gamma_{ln}{}^{j)}, \label{Om3}\\
\Omega_k{}^{ijn} &=& {\frac {\partial {\cal L}}{\partial \partial_{(n}\Gamma_{ij)}{}^k}}.\label{Om4}
\end{eqnarray}
If the action is invariant under general coordinate transformations, $\delta I = 0$, in view of the arbitrariness of the function $\xi^i$ and its derivatives, we find the set of four Noether identities:
\begin{equation}\label{NoeG}
\Omega_k = 0,\quad \Omega_k{}^i = 0,\quad \Omega_k{}^{ij} = 0, \quad \Omega_k{}^{ijn} = 0.
\end{equation}

General coordinate invariance is a natural consequence of the fact that the action (\ref{action}) and the Lagrangian ${\cal L}$ are constructed only from covariant objects. Namely, ${\cal L} = {\cal L}(\psi^A, \nabla_i\psi^A, g_{ij}, R_{kli}{}^j, N_{kj}{}^i)$ is a function of the metric, the curvature (\ref{curv}), the torsion (\ref{tors}), the matter fields, and their {\it covariant derivatives} 
\begin{equation}\label{Dpsi}
\nabla_k\psi^A = \partial_k\psi^A -\Gamma_{ki}{}^j\,(\sigma^A{}_B)_j{}^i\,\psi^B.
\end{equation}
Denoting 
\begin{eqnarray}
\rho^{ijk}{}_l := {\frac {\partial {\cal L}}{\partial R_{ijk}{}^l}},\qquad
\mu^{ij}{}_k := {\frac {\partial {\cal L}}{\partial N_{ij}{}^k}},\label{dLRN}
\end{eqnarray}
we find for the derivatives of the Lagrangian
\begin{eqnarray}
{\frac {\partial {\cal L}}{\partial \Gamma_{ij}{}^k}} &=& -\,{\frac {\partial {\cal L}} {\partial \nabla_i\psi^A}}(\sigma^A{}_B)_k{}^j\,\psi^B - \mu^{ij}{}_k\nonumber\\
&& + 2\rho^{inl}{}_k\Gamma_{nl}{}^j + 2\rho^{nij}{}_l\Gamma_{nk}{}^l,\label{dLG}\\
{\frac {\partial {\cal L}}{\partial \partial_i\Gamma_{jk}{}^l}} &=& 2\rho^{ijk}{}_l,\label{dLdG}\\ 
{\frac {\partial {\cal L}}{\partial \partial_kg_{ij}}} &=& {\frac 12}\left(\mu^{(ki)j} + \mu^{(kj)i} - \mu^{(ij)k}\right).\label{dLdm}
\end{eqnarray}
As a result, we straightforwardly verify that $\Omega_k{}^{ij} = 0$ and $\Omega_k{}^{ijn} = 0$ are indeed satisfied identically. 

Using (\ref{dLG}) and (\ref{dLdG}), we then recast the two remaining Noether identities (\ref{Om1}) and (\ref{Om2}) into
\begin{eqnarray}
\hspace{-0.2cm}\Omega_k &=& {\frac {\delta {\cal L}}{\delta g_{ij}}}\,\partial_kg_{ij} + {\frac {\delta {\cal L}}{\delta\psi^A}}\,\partial_k\psi^A + \partial_i\!\left(\!{\frac {\partial {\cal L}}{\partial\nabla_i\psi^A}}\nabla_k\psi^A - \delta^i_k{\cal L} \!\right)\!\nonumber\\
&& +\,\widehat{\nabla}{}_j\!\left(\!{\frac {\partial {\cal L}}{\partial\nabla_j\psi^A}} \,(\sigma^A{}_B)_m{}^n\,\psi^B\!\right)\!\Gamma_{kn}{}^m + \rho^{iln}{}_m\partial_kR_{iln}{}^m\nonumber\\
&& + \,{\frac {\partial {\cal L}}{\partial\nabla_l\psi^A}}\,(\sigma^A{}_B)_m{}^n \,\psi^B\,R_{lkn}{}^m + \mu^{ln}{}_m\partial_kN_{ln}{}^m \nonumber\\
&& +\,{\frac 12}\check{\nabla}_i\left(\mu^{(im)n} + \mu^{(in)m} - \mu^{(mn)i}\right)\partial_k g_{mn},\label{Om1a}\\
\hspace{-0.2cm}\Omega_k{}^i &=& 2{\frac {\delta {\cal L}}{\delta g_{ij}}}\,g_{kj} + {\frac {\delta {\cal L}}{\delta\psi^A}}\,(\sigma^A{}_B)_k{}^i\,\psi^B + {\frac {\partial {\cal L}}{\partial\nabla_i\psi^A}}\nabla_k\psi^A 
\nonumber\\ 
&& - \delta^i_k{\cal L} + \widehat{\nabla}{}_j\left({\frac {\partial {\cal L}}{\partial\nabla_j\psi^A}}(\sigma^A{}_B)_k{}^i\psi^B\right)  - \mu^{ln}{}_kN_{ln}{}^i  \nonumber\\
&& + \mu^{il}{}_nN_{kl}{}^n+ \mu^{li}{}_nN_{lk}{}^n+ \,2\rho^{iln}{}_mR_{kln}{}^m \nonumber\\
&& + \rho^{lni}{}_mR_{lnk}{}^m - \rho^{lnm}{}_kR_{lnm}{}^i \nonumber\\
&& +\,\check{\nabla}_n\left(\mu^{(ni)j} + \mu^{(nj)i} - \mu^{(ij)n}\right)g_{jk}= 0.\label{Om2a}
\end{eqnarray}
Here we introduced the covariant derivative for an arbitrary tensor density ${\cal A}^n{}_{i\dots}{}^{j\dots}$
\begin{equation}
\widehat{\nabla}{}_n{\cal A}^n{}_{i\dots}{}^{j\dots} = \partial_n{\cal A}^n{}_{i\dots}{}^{j\dots} + \Gamma_{nl}{}^j{\cal A}^n{}_{i\dots}{}^{l\dots} - \Gamma_{ni}{}^l {\cal A}^n{}_{l\dots}{}^{j\dots},\label{dA}
\end{equation}
which produces a tensor density of the same weight. We denote a similar differential operation constructed with the help of the Riemannian connection by
\begin{equation}
\check{\nabla}{}_n{\cal A}^n{}_{i\dots}{}^{j\dots} = \partial_n{\cal A}^n{}_{i\dots}{}^{j\dots} + \widetilde{\Gamma}_{nl}{}^j{\cal A}^n{}_{i\dots}{}^{l\dots} - \widetilde{\Gamma}_{ni}{}^l
{\cal A}^n{}_{l\dots}{}^{j\dots},\label{dAc}
\end{equation}
In particular, notice that the variational derivative (\ref{var}) w.r.t.\ the matter fields can be identically rewritten as
\begin{equation}\label{covar}
{\frac {\delta {\cal L}}{\delta\psi^A}} = {\frac {\partial {\cal L}}{\partial\psi^A}} - \widehat{\nabla}{}_j\left({\frac {\partial {\cal L}}{\partial\nabla_j\psi^A}}\right),
\end{equation}
and turns out to be a covariant tensor density. It is also worthwhile to note, that the variational derivative w.r.t.\ the metric is explicitly a covariant density. This follows from the fact that the Lagrangian depends on $g_{ij}$ not only directly, but also through the objects $Q_{kij}$ and $N_{ki}{}^j$. Taking this into account, we find
\begin{eqnarray}
{\frac {\delta {\cal L}}{\delta g_{ij}}} &=& {\frac {d {\cal L}}{d g_{ij}}} - \partial_n\left(
{\frac {\partial {\cal L}}{\partial \partial_ng_{ij}}}\right) \nonumber\\
&=& {\frac {\partial {\cal L}}{\partial g_{ij}}}- {\frac 12}\check{\nabla}_n\left(\mu^{(ni)j} + \mu^{(nj)i} - \mu^{(ij)n}\right).\label{dLgij}
\end{eqnarray}

The Noether identity (\ref{Om2a}) is a covariant relation. In contrast, (\ref{Om1a}) is apparently non-covariant. However, this can be easily repaired by replacing $\Omega_k = 0$ with an equivalent covariant Noether identity: $\overline{\Omega}{}_k = \Omega_k -  \Gamma_{kn}{}^m\Omega_m{}^n = 0$. Explicitly, we find
\begin{eqnarray}
\overline{\Omega}{}_k &=& {\frac {\delta {\cal L}}{\delta\psi^A}}\,\nabla_k\psi^A + \widehat{\nabla}{}_i\!\left(\!{\frac {\partial {\cal L}}{\partial\nabla_i\psi^A}} \,\nabla_k\psi^A - \delta^i_k{\cal L} \!\right)\nonumber\\ 
&&  - \left({\frac {\partial {\cal L}}{\partial\nabla_i\psi^A}} \,\nabla_l\psi^A - \delta^i_l{\cal L}\right) T_{ki}{}^l + \mu^{ln}{}_m\nabla_kN_{ln}{}^m  \nonumber\\
&& +\,{\frac {\partial {\cal L}}{\partial\nabla_l\psi^A}}\,(\sigma^A{}_B)_m{}^n \,\psi^B\,R_{lkn}{}^m + \rho^{iln}{}_m\nabla_kR_{iln}{}^m   \nonumber\\ 
&& + \left[ - {\frac {\delta {\cal L}}{\delta g_{ij}}} - {\frac 12}\check{\nabla}_n\left(\mu^{(ni)j} + \mu^{(nj)i} - \mu^{(ij)n}\right)\right]Q_{kij} \nonumber \\ &= & 0.\label{Om1b}
\end{eqnarray}

On-shell, i.e., assuming that the matter fields satisfy the field equations ${\delta {\cal L}}/{\delta\psi^A} = 0$, the Noether identities (\ref{Om1b}) and (\ref{Om2a}) reduce to the {\it conservation laws} for the energy-momentum and hypermomentum, respectively.

Equation (\ref{Om2a}) contains a relation between the canonical and the metrical energy-momentum tensor, and the conservation law of the hypermomentum. In the next section we turn to the discussion of models with general nonminimal coupling.

\section{Conservation laws in models with nonminimal coupling}\label{conservation_sec}

The results obtained in the previous section are applicable to {\it any} theory in which the Lagrangian depends arbitrarily on the matter fields and the gravitational field strengths. Now we specialize to the class of models described by an interaction Lagrangian of the form
\begin{equation}
L = F(g_{ij},R_{kli}{}^j, N_{kl}{}^i)\,L_{\rm mat}(\psi^A, \nabla_i\psi^A).\label{LFL}
\end{equation}
Here $L_{\rm mat}(\psi^A, \nabla_i\psi^A)$ is the ordinary matter Lagrangian. We call  $F = F(g_{ij},R_{kli}{}^j, N_{kl}{}^i)$ the coupling function and assume that it can depend arbitrarily on its arguments, i.e., on all covariant gravitational field variables of MAG. When $F = 1$ we recover the minimal coupling case. 

\subsection{Identities for the nonminimal coupling function}\label{Fidentities_subsec}

As a preliminary step, let us derive identities which are satisfied for the nonminimal coupling function $F = F(g_{ij},R_{kli}{}^j, N_{kl}{}^i)$. For this, we apply the above Lagrange-Noether machinery to the auxiliary Lagrangian density ${\cal L}_0 = \sqrt{-g}\,F$. This quantity does not depend on the matter fields, and both (\ref{Om2a}) and (\ref{Om1b}) are considerably simplified. In particular, we have
\begin{equation}
{\frac {\delta {\cal L}_0}{\delta g_{ij}}} = \sqrt{-g}\left({\frac 12}Fg^{ij} + F^{ij} \right),\qquad F^{ij} := {\frac {\delta F}{\delta g_{ij}}}.\label{dFg}
\end{equation}
Then we immediately see that (\ref{Om2a}) and (\ref{Om1b}) reduce to
\begin{eqnarray}
\nabla_k F &=& \left[-F^{ij} - {\frac 12}\widetilde{\nabla}_n\left({\stackrel 0 \mu}{}^{(ni)j} + {\stackrel 0 \mu}{}^{(nj)i} - {\stackrel 0 \mu}{}^{(ij)n}\right)\right]Q_{kij} \nonumber\\
&& + \,{\stackrel 0 \rho}{}^{iln}{}_m\nabla_kR_{iln}{}^m + {\stackrel 0 \mu}{}^{ln}{}_m\nabla_kN_{ln}{}^m,\label{F1}\\
2F_k{}^i &=& - \,2{\stackrel 0 \rho}{}^{iln}{}_mR_{kln}{}^m - {\stackrel 0 \rho}{}^{lni}{}_m R_{lnk}{}^m + {\stackrel 0 \rho}{}^{lnm}{}_kR_{lnm}{}^i\nonumber\\
&& - \,{\stackrel 0 \mu}{}^{ln}{}_kN_{ln}{}^i  + {\stackrel 0 \mu}{}^{il}{}_nN_{kl}{}^n + {\stackrel 0 \mu}{}^{li}{}_nN_{lk}{}^n\nonumber\\
&& +\,\widetilde{\nabla}_n\left({\stackrel 0 \mu}{}^{(ni)j} + {\stackrel 0 \mu}{}^{(nj)i} - {\stackrel 0 \mu}{}^{(ij)n}\right)g_{jk}.\label{F2}
\end{eqnarray}
Here we denoted
\begin{equation}\label{dFRT}
{\stackrel 0 \rho}{}^{ijk}{}_l := {\frac {\partial F}{\partial R_{ijk}{}^l}},\qquad
{\stackrel 0 \mu}{}^{ij}{}_k := {\frac {\partial F}{\partial N_{ij}{}^k}}.
\end{equation}
Notice that for any tensor density ${\cal A}^n{}_{i\dots}{}^{j\dots} = \sqrt{-g}A^n{}_{i\dots}{}^{j\dots}$ we have
\begin{eqnarray}\label{nablatilde}
\check{\nabla}_n{\cal A}^n{}_{i\dots}{}^{j\dots} = \sqrt{-g}\,\widetilde{\nabla}{}_nA^n{}_{i\dots}{}^{j\dots}.
\end{eqnarray}
Similarly for the non-Riemannian derivatives (\ref{dA}) of tensor densities we find
\begin{eqnarray}\label{nablastar}
\widehat{\nabla}_n{\cal A}^n{}_{i\dots}{}^{j\dots} = \sqrt{-g}\,{\stackrel * \nabla}{}_nA^n{}_{i\dots}{}^{j\dots},
\end{eqnarray}
where the so-called modified covariant derivative is defined as
\begin{equation}
{\stackrel * \nabla}{}_i = \nabla_i + N_{ki}{}^k.\label{dstar}
\end{equation}
The identity (\ref{F1}) is naturally interpreted as a generally covariant generalization of the chain rule for the total derivative of a function of several variables. This becomes obvious when we notice that (\ref{dLgij}) implies
\begin{eqnarray}
&&\left[-F^{ij} - {\frac 12}\widetilde{\nabla}_n\left({\stackrel 0 \mu}{}^{(ni)j} + {\stackrel 0 \mu}{}^{(nj)i} - {\stackrel 0 \mu}{}^{(ij)n}\right)\right]Q_{kij} \nonumber \\
&&= {\frac {\partial F} {\partial g_{ij}}}\,\nabla_kg_{ij}.\label{difFg}
\end{eqnarray}
It should be stressed that (\ref{F1}) and (\ref{F2}) are true identities, they are satisfied for any function $F(g_{ij}, R_{kli}{}^j, N_{kl}{}^i)$ irrespectively of the field equations that can be derived from the corresponding action. 

\subsection{Conservation laws}\label{Conslaws_subsec}

Now we are in a position to derive the conservation laws for the general nonminimal coupling model (\ref{ansatz_lagrangian_model_3}), and thus we have to consider the Lagrangian density
\begin{equation}
{\cal L} = \sqrt{-g}FL_{\rm mat}.\label{Lnon}
\end{equation}
As before, $F = F(g_{ij}, R_{kli}{}^j, N_{kl}{}^i)$ is an arbitrary function of its arguments, whereas the matter Lagrangian $L_{\rm mat} = L_{\rm mat}(\psi^A, \nabla_i\psi^A, g_{ij})$ has the usual form established from the minimal coupling principle. 

In a standard way, matter is characterized by the canonical energy-momentum tensor
\begin{equation}
\Sigma_k{}^i = {\frac {\partial {L_{\rm mat}}}{\partial\nabla_i\psi^A}} \,\nabla_k\psi^A - \delta^i_kL_{\rm mat},\label{emcan}
\end{equation}
the canonical hypermomentum tensor,
\begin{equation}
\Delta^n{}_k{}^i = -\,{\frac {\partial {L_{\rm mat}}}{\partial\nabla_i\psi^A}} \,(\sigma^A{}_B)_k{}^n \psi^B,\label{spin}
\end{equation}
and the metrical energy-momentum tensor
\begin{equation}\label{emmet}
t_{ij} = {\frac 2{\sqrt{-g}}}\,{\frac {\delta {(\sqrt{-g}L_{\rm mat})}}{\delta g^{ij}}}.
\end{equation}
The usual spin arises as the antisymmetric part of the hypermomentum,
\begin{equation}
\tau_{ij}{}^k = \Delta_{[ij]}{}^k,\label{spindef}
\end{equation}
whereas the trace $\Delta^k = \Delta^i{}_i{}^k$ is the dilation current. The symmetric traceless part describes the proper hypermomentum \cite{Hehl:1995}.

In view of the product structure of the Lagrangian (\ref{Lnon}), the derivatives are easily evaluated, and the conservation laws (\ref{Om2a}) and (\ref{Om1b}) reduce to
\begin{eqnarray}
&& -\,Ft_k{}^i - {\stackrel * \nabla}{}_n\left(F\Delta^i{}_k{}^n\right) + F\Sigma_k{}^i + \Bigl[2F_k{}^i \nonumber\\
&& +\,2{\stackrel 0 \rho}{}^{iln}{}_mR_{kln}{}^m + {\stackrel 0 \rho}{}^{lni}{}_m R_{lnk}{}^m - {\stackrel 0 \rho}{}^{lnm}{}_kR_{lnm}{}^i \nonumber\\ 
&& +\,{\stackrel 0 \mu}{}^{ln}{}_kN_{ln}{}^i  - {\stackrel 0 \mu}{}^{il}{}_nN_{kl}{}^n - {\stackrel 0 \mu}{}^{li}{}_nN_{lk}{}^n  \nonumber\\ 
&& -\,\widetilde{\nabla}_n\left({\stackrel 0 \mu}{}^{(ni)j} + {\stackrel 0 \mu}{}^{(nj)i} - {\stackrel 0 \mu}{}^{(ij)n}\right)g_{jk}\Bigr]L_{\rm mat} = 0, \label{cons1a}\\
&& {\stackrel * \nabla}{}_i\left(F\Sigma_k{}^i\right)  + F\Bigl[ - \Sigma_l{}^i T_{ki}{}^l + \Delta^m{}_n{}^l R_{klm}{}^n +  {\frac 12}t^{ij}Q_{kij}\Bigr] \nonumber\\
&& + \Bigl\{{\stackrel 0 \rho}{}^{iln}{}_m\nabla_kR_{iln}{}^m + {\stackrel 0 \mu}{}^{ln}{}_m\nabla_kN_{ln}{}^m + \Bigl[-F^{ij} \nonumber\\
&&- {\frac 12}\widetilde{\nabla}_n\!\Bigl({\stackrel 0 \mu}{}^{(ni)j} + {\stackrel 0 \mu}{}^{(nj)i} - {\stackrel 0 \mu}{}^{(ij)n}\Bigr)\Bigr]Q_{kij}\Bigr\}L_{\rm mat} = 0.\label{cons2a}
\end{eqnarray}
After we take into account the identities (\ref{F1}) and (\ref{F2}), the conservation laws (\ref{cons1a}) and (\ref{cons2a}) are brought to the final form:
\begin{eqnarray}
F\Sigma_k{}^i &=& Ft_k{}^i + {\stackrel * \nabla}{}_n\left(F\Delta^i{}_k{}^n\right),\label{cons1b}\\
{\stackrel * \nabla}{}_i\left(F\Sigma_k{}^i\right) &=& F \left( \Sigma_l{}^i T_{ki}{}^l - \Delta^m{}_n{}^l R_{klm}{}^n  - {\frac 12}t^{ij}Q_{kij} \right) \nonumber \\
&&- L_{\rm mat}\nabla_kF.\label{cons2b}
\end{eqnarray}
Lowering the index in (\ref{cons1b}) and antisymmetrizing, we derive the conservation law for the spin
\begin{equation}\label{skewS}
F\Sigma_{[ij]} + {\stackrel * \nabla}{}_n\left(F\tau_{ij}{}^n\right) + Q_{nl[i}\Delta^l{}_{j]}{}^n = 0.
\end{equation}
This is a generalization of the usual conservation law of the total angular momentum for the case of nonminimal coupling. 

\subsection{Riemannian limit}\label{Riemannian_subsec}

Our results contain the Riemannian theory as a special case. Suppose that the torsion and the nonmetricity are absent $T_{ij}{}^k = 0$, $Q_{kij} = 0$, hence $N_{ij}{}^k = 0$. Then for usual matter without microstructure (i.e.\ matter with $\Delta^m{}_n{}^i = 0$) the canonical and the metrical energy-momentum tensors coincide, $\Sigma_k{}^i = t_k{}^i$. As a result, the conservation law (\ref{cons2b}) reduces to
\begin{equation}
\nabla_it_k{}^i = {\frac 1F}\left(- L_{\rm mat}\delta_k^i -t_k{}^i\right)\nabla_iF.\label{consF}
\end{equation}
This conservation law for the general nonminimal coupling model was derived earlier in \cite{Puetzfeld:Obukhov:2013} without using the Noether theorem, directly from the field equations\footnote{Notice a different conventional sign, as compared to our previous work \cite{Puetzfeld:Obukhov:2013}.}. The old result established the conservation law for the case in which $F = F(R_{ijk}{}^l)$ depends arbitrarily on the components of the curvature tensor, correcting some erroneous derivations in the literature, see \cite{Puetzfeld:Obukhov:2013} for details. 

Quite remarkably, (\ref{consF}) generalizes the earlier result to the case in which the nonminimal coupling function $F$ is a general scalar function of the curvature tensor. 

\section{Field dynamics in metric-affine gravity}\label{MAG}

The explicit form of the dynamical equations of the gravitational field is irrelevant for the conservation laws that we derived in the previous sections solely on the basis of the Noether theorem. For completeness, however, we discuss now the field equations of a general metric-affine theory of gravity. The standard understanding of MAG is its interpretation as a gauge theory based on the general affine group $GA(4,R)$, which is a semidirect product of the general linear group $GL(4,R)$, and the group of local translations \cite{Hehl:1995}. The corresponding gauge-theoretic formalism generalizes the approach of Sciama and Kibble \cite{Sciama:1962,Kibble:1961}; for more details about gauge gravity theories, see \cite{Blagojevic:2002,Hehl:2013}. In the standard formulation of MAG as a gauge theory \cite{Hehl:1995}, the gravitational gauge potentials are identified with the metric, coframe, and the linear connection. The corresponding gravitational field strengths are then the nonmetricity, the torsion, and the curvature, respectively. 

In the present paper we use an alternative formulation of MAG in which gravity is described by a different set of fundamental field variables, i.e.\ the independent metric $g_{ij}$ and connection $\Gamma_{ki}{}^j$. For the relevant literature, see \cite{Hehl:1976a,Hehl:1976b,Hehl:1976c,Hehl:1978,Hehl:1981,Obukhov:1982,Vassiliev:2005,Sotiriou:2007,Sotiriou:2010,Vitagliano:2011}, for example. It is worthwhile to compare the field equations in the different formalisms of MAG, and in particular, it is necessary to clarify the role and place of the {\it canonical} energy-momentum tensor as a source of the gravitational field. Since one does not have the coframe (tetrad) among the fundamental variables, the corresponding field equation is absent. Here we demonstrate that one can always rearrange the field equations of MAG in such a way that the canonical energy-momentum tensor is recovered as one of the sources of the gravitational field. 

Let us consider the total Lagrangian density of coupled gravitational and matter fields:
\begin{equation}
{\cal L} = {\cal V}(g_{ij}, R_{ijk}{}^l, N_{ki}{}^j) + {\cal L}_{\rm mat}(g_{ij}, \psi^A, \nabla_i\psi^A).\label{Ltot}
\end{equation}
Then, from the variation of the action with respect to the metric $g_{ij}$ and the connection $\Gamma_{ki}{}^j$, we derive the field equations:
\begin{eqnarray}
2{\frac {\delta{\cal V}}{\delta g_{ij}}} &=& \hat{t}^{ij},\label{0th}\\
\hat{\nabla}_l{\cal H}^{kli}{}_j + {\frac 12}T_{mn}{}^k{\cal H}^{mni}{}_j - {\cal E}^{ki}{}_j &=& \hat{\Delta}^i{}_j{}^k.\label{2nd}  
\end{eqnarray}
Here, we introduced the generalized gravitational field momentum density
\begin{equation}
{\cal H}^{kli}{}_j = -\,2{\frac {\partial{\cal V}}{\partial R_{kli}{}^j}},\label{HH}
\end{equation}
and the gravitational hypermomentum density
\begin{equation}
{\cal E}^{ki}{}_j = -\,{\frac {\partial{\cal V}}{\partial N_{ki}{}^j}}.\label{EN}
\end{equation}
The right-hand sides of (\ref{0th}) and (\ref{2nd}) are the metric energy-momentum density and the hypermomentum density of matter, respectively,
\begin{equation}
\hat{t}_{ij} = 2{\frac {\partial{\cal L}_{\rm mat}}{\partial g^{ij}}},\quad \hat{\Delta}^i{}_j{}^k = {\frac {\partial{\cal L}_{\rm mat}}{\partial \Gamma_{ki}{}^j}} =  - {\frac {\partial {{\cal L}_{\rm mat}}}{\partial\nabla_i\psi^A}} \,(\sigma^A{}_B)_k{}^n \psi^B.\label{tD}
\end{equation}
Here we assume minimal coupling of matter and gravity. For ${\cal L}_{\rm mat} = \sqrt{-g}L_{\rm mat}$, comparing with (\ref{emmet}) and (\ref{spin}), we immediately find $\hat{t}_{ij} = \sqrt{-g}\,t_{ij}$ and $\hat{\Delta}^i{}_j{}^k = \sqrt{-g}\,{\Delta}^i{}_j{}^k$. 

To reveal the role of the canonical energy-momentum tensor, we use the Noether identities arising when the theory is invariant under general coordinate transformations. Then, applying (\ref{Om2a}) to the {\it total Lagrangian} (\ref{Ltot}), we find the Noether identity
\begin{eqnarray}
\Omega_k{}^i &=& 2{\frac {\delta {\cal L}}{\delta g_{ij}}}\,g_{kj} + {\frac {\delta {\cal L}}{\delta \psi^A}}(\sigma^A{}_B)_k{}^i\psi^B  + \hat{\Sigma}_k{}^i - \delta_k^i{\cal V} \nonumber\\
&& + {\cal E}^{ln}{}_kN_{ln}{}^i - {\cal E}^{il}{}_nN_{kl}{}^n - {\cal E}^{li}{}_nN_{lk}{}^n  -\,\hat{\nabla}_j\hat{\Delta}^i{}_k{}^j\nonumber\\
&& - \check{\nabla}_n\left({\cal E}^{(ni)j} + {\cal E}^{(nj)i} - {\cal E}^{(ij)n}\right)g_{jk} -\,R_{kln}{}^m{\cal H}^{iln}{}_m \nonumber\\
&&+ {\frac 12}R_{lnm}{}^i{\cal H}^{lnm}{}_k - {\frac 12}R_{lnk}{}^m{\cal H}^{lni}{}_m  = 0.\label{Om2M}
\end{eqnarray}
Here the canonical energy-momentum density of matter reads, cf.\ with (\ref{emcan}),
\begin{equation}
\hat{\Sigma}_k{}^i = {\frac {\partial {{\cal L}_{\rm mat}}}{\partial\nabla_i\psi^A}}\,\nabla_k\psi^A - \delta^i_k{\cal L}_{\rm mat}.\label{canD}
\end{equation}

Suppose, as it is assumed in the gauge-theoretic approach to MAG \cite{Hehl:1995}, that the gravitational Lagrangian depends on the post-Riemannian geometric variables only via the torsion and the nonmetricity. Then, using (\ref{torsN})-(\ref{nonmetN}), we find
\begin{equation}
{\cal E}^{ki}{}_j = - {\cal H}^{ki}{}_j - {\cal M}^{ki}{}_j.\label{EHM}
\end{equation}
Here we introduced 
\begin{equation}
{\cal H}^{ki}{}_j = -\,{\frac {\partial{\cal V}}{\partial T_{ki}{}^j}},\qquad
{\cal M}^{kij} = -\,{\frac {\partial{\cal V}}{\partial Q_{kij}}}.\label{HM}
\end{equation}
Substituting (\ref{EHM}) into (\ref{Om2M}), after some long algebra, we find on ``mass-shell'' (i.e., when the field equations are satisfied $\delta{\cal L}/\delta g_{ij} = 0$ and $\delta{\cal L}/\delta\psi^A = 0$)\footnote{The identity $\hat{\nabla}_l\Big(\hat{\nabla}_n{\cal H}^{lni}{}_k + {\frac 12}T_{mn}{}^l{\cal H}^{mni}{}_k\Big) \equiv  {\frac 12}R_{lnm}{}^i{\cal H}^{lnm}{}_k - {\frac 12}R_{lnk}{}^m{\cal H}^{lni}{}_m$ can be straightforwardly verified.}:
\begin{eqnarray}
\Omega_k{}^i &=&  \hat{\Sigma}_k{}^i - {\cal E}_k{}^i  + {\frac 12}T_{mn}{}^i{\cal H}^{mn}{}_k +\,\hat{\nabla}_l\Big(\hat{\nabla}_n{\cal H}^{lni}{}_k \nonumber\\
&& + {\frac 12}T_{mn}{}^l{\cal H}^{mni}{}_k + {\cal M}^{li}{}_k - \hat{\Delta}^i{}_k{}^l\Big) = 0.\label{Om2F}
\end{eqnarray}
Here we introduced
\begin{equation}
{\cal E}_k{}^i = \delta_k^i{\cal V} + {\frac 12}Q_{kln}{\cal M}^{iln} + T_{kl}{}^n{\cal H}^{il}{}_n + R_{kln}{}^m{\cal H}^{iln}{}_m.\label{Eg}
\end{equation}
Finally, after inserting (\ref{2nd}) into (\ref{Om2F}), we recast the latter into a form of the field equation
\begin{equation}\label{1st}
\hat{\nabla}_n{\cal H}^{in}{}_k + {\frac 12}T_{mn}{}^i{\cal H}^{mn}{}_k - {\cal E}_k{}^i = - \hat{\Sigma}_k{}^i.
\end{equation}
The system of the three field equations (\ref{0th}), (\ref{2nd}), and (\ref{1st}) perfectly reproduces the gauge-theoretic structure of MAG (notice some different sign conventions in \cite{Hehl:1995}). 

It is worthwhile to note that ``on-shell'' (when the gravitational and the matter fields satisfy the Lagrange-Euler equations $\delta{\cal L}/\delta g_{ij} = 0, \delta{\cal L}/\delta \Gamma_{ij}{}^k = 0$, and $\delta{\cal L}/\delta\psi^A = 0$) the Noether identities (\ref{Om1}) and (\ref{Om2}) can be rewritten as the ordinary conservation laws
\begin{equation}\label{dTdS}
\partial_i{\mathfrak T}_k{}^i = 0,\qquad {\mathfrak T}_k{}^i + \partial_j{\mathfrak S}^i{}_k{}^j = 0.
\end{equation}
The total energy-momentum and hypermomentum {\it pseudotensors} of the interacting gravitational and matter fields 
\begin{eqnarray}
{\mathfrak T}_k{}^i &=& {\frac {\partial {\cal L}}{\partial\partial_i g_{mn}}}\partial_kg_{mn} + {\frac {\partial {\cal L}}{\partial\partial_i \Gamma_{ln}{}^m}}\partial_k\Gamma_{ln}{}^m \nonumber \\
&&+ {\frac {\partial {\cal L}}{\partial\partial_i \psi^A}}\partial_k\psi^A  - \delta_k^i{\cal L}, \label{TP}\\
{\mathfrak S}^i{}_k{}^j &=& {\frac {\partial {\cal L}}{\partial\partial_j \Gamma_{ln}{}^m}}\left(\delta^i_l\Gamma_{kn}{}^m + \delta^i_n\Gamma_{lk}{}^m - \delta^m_k\Gamma_{ln}{}^i\right)\nonumber\\
&& +\,2{\frac {\partial {\cal L}}{\partial\partial_j g_{il}}} g_{kl}+ {\frac {\partial {\cal L}}{\partial\partial_j \psi^A}}(\sigma^A{}_B)_k{}^i\psi^B,\label{SP}
\end{eqnarray}
generalize the well-known general-relativistic energy-momentum pseudotensor \cite{Obukhov:2007}. It is interesting that we can recast the second equation (\ref{dTdS}) into a conservation law of the ``orbital $+$ intrinsic'' hypermomentum:
\begin{equation}
\partial_j{\mathfrak J}^i{}_k{}^j = 0,\qquad {\mathfrak J}^i{}_k{}^j = x^i{\mathfrak T}_k{}^j
+ {\mathfrak S}^i{}_k{}^j.\label{dJ}
\end{equation}

\section{General relativity: nonminimal coupling of matter with spin}\label{quadrupole}

In previous studies of general-relativistic models with nonminimal coupling \cite{Puetzfeld:Obukhov:2013}, the matter was assumed to be spinless. The present formalism is suitable for the description of matter with (intrinsic) spin. Let us consider the Lagrangian density ${\cal L} = \sqrt{-g}\,FL_{\rm mat}$, where the coupling function depends arbitrarily on the Riemann curvature $F = F(R_{klm}{}^n)$. We can then use the general metric-affine scheme, with independent metric and connection, and recover the Riemannian geometry by imposing the constraint $N_{ki}{}^j = 0$. The constraint is taken into account by means of the Lagrange multiplier $\phi^{ki}{}_j$ so that the Lagrangian reads 
\begin{equation}
{\cal L} = \sqrt{-g}(FL_{\rm mat} + \phi^{ki}{}_jN_{ki}{}^j).\label{Lc}
\end{equation}
The derivatives (\ref{dLRN}) are straightforwardly computed:
\begin{eqnarray}
\rho^{ijk}{}_l &=& \sqrt{-g}\,{\stackrel 0 \rho}{}^{ijk}{}_l,\qquad \mu^{ij}{}_k = \sqrt{-g}\,\phi^{ij}{}_k.\label{snrm}
\end{eqnarray}
The constraint $N_{ki}{}^j = 0$ eliminates the connection $\Gamma_{ki}{}^j$ as an independent variable, and hence we have to take into account the corresponding field equations
\begin{equation}
{\frac {\delta{\cal L}}{\delta\Gamma_{ki}{}^j}} = F\Delta^i{}_j{}^k - \phi^{ki}{}_j - 2\nabla_n\left({\stackrel 0 \rho}{}^{nki}{}_jL_{\rm mat}\right) = 0.\label{dLdGijk}
\end{equation}
This allows us to find explicitly
\begin{equation}
\mu^{ij}{}_k = \sqrt{-g}\left[F\Delta^j{}_k{}^i - 2\nabla_n\left(L_{\rm mat}{\stackrel 0 \rho}{}^{nij}{}_k\right)\right],\label{mu}
\end{equation}
and consequently
\begin{eqnarray}
\mu^{(ni)j} + \mu^{(nj)i} - \mu^{(ij)n} &&= \sqrt{-g}\Big[2F\tau^{n(ij)} + F\Delta^{(ij)n} \nonumber \\
&&- 4\nabla_m\left({\stackrel 0 \rho}{}^{m(i|n|j)}L_{\rm mat}\right)\Big].\label{mmm}
\end{eqnarray}
Note that by construction we have the symmetry properties 
\begin{equation}
{\stackrel 0 \rho}{}^{ijkl} = {\stackrel 0 \rho}{}^{[ij]kl} 
= {\stackrel 0 \rho}{}^{ij[kl]} = {\stackrel 0 \rho}{}^{klij}.\label{Isym} 
\end{equation}
We will repeatedly use these symmetries in the subsequent computations. 

As a first step in the derivation of the conservation laws, we notice that the Noether identities for the coupling function $F$ are still valid. In the Riemannian case with $N_{ki}{}^j = 0$, (\ref{F1}) and (\ref{F2}) are reduced to
\begin{eqnarray}
\nabla_kF &=& {\stackrel 0 \rho}{}^{iln}{}_m\nabla_kR_{iln}{}^m,\label{F1a}\\
{\frac {\partial F}{\partial g_{ij}}}\,g_{kj} &=& -\,{\stackrel 0 \rho}{}_{lmn}{}^iR^{lmn}{}_k.\label{F2a}
\end{eqnarray}
It is worthwhile to notice that the last identity implies the symmetry property ${\stackrel 0 \rho}{}_{lmn}{}^iR^{lmn}{}_k = {\stackrel 0 \rho}{}^{lmn}{}_kR_{lmn}{}^i$. 

We are now in a position to derive the main result of this section. Using the Riemannian geometry constraint $N_{ki}{}^j = 0$, and substituting  (\ref{mmm}), (\ref{F1a}), and (\ref{F2a}) into the Noether identities (\ref{Om2a}) and (\ref{Om1b}), we find on-shell the conservation laws:
\begin{eqnarray}
&&\Pi_k{}^i = F\Sigma_k{}^i + {\nabla}{}_n\left[F\left(\tau^{ni}{}_k + \tau^n{}_k{}^i - \tau^i{}_k{}^n\right)\right]\nonumber\\
&& - \,4\nabla_n\nabla_m\left(L_{\rm mat}{\stackrel 0 \rho}{}^{nim}{}_k\right) + 2L_{\rm mat}{\stackrel 0 \rho}{}_{lmn}{}^iR^{lmn}{}_k,\label{q1a}\\
&&{\nabla}{}_i(F\Sigma_k{}^i) = -\,F\tau^m{}_n{}^l R_{klm}{}^n - L_{\rm mat}\nabla_kF.\label{q2a}
\end{eqnarray}
Here, we denoted 
\begin{equation}\label{Pi}
\Pi_{ij} = {\frac 2{\sqrt{-g}}}\,{\frac {\delta {\cal L}}{\delta g^{ij}}}.
\end{equation}
This tensor describes the right-hand side of the gravitational field equation, i.e., it is the generalized source. In the special case of minimal coupling ($F = 1$), it reduces to the metrical energy-momentum tensor, $\Pi_{ij} = t_{ij}$. 

The covariant divergence of this tensor vanishes as a consequence of the conservation laws (\ref{q1a}) and (\ref{q2a}):
\begin{equation}
\nabla_i\Pi_k{}^i = 0.\label{dPi}
\end{equation}
The proof is straightforward but somewhat lengthy. At first we notice that in view of the skew symmetry we have $\nabla_i\nabla_n\left[F\left(\tau^{ni}{}_k + \tau^n{}_k{}^i - \tau^i{}_k{}^n\right)\right] = \nabla_{[i}\nabla_{n]}$ $\left[F\left(\tau^{ni}{}_k + \tau^n{}_k{}^i - \tau^i{}_k{}^n\right)\right]$, and because of (\ref{Isym}) similarly $\nabla_i\nabla_n\nabla_m(L_{\rm mat}{\stackrel 0 \rho}{}^{nim}{}_k) = \nabla_{[i}\nabla_{n]}\nabla_m(L_{\rm mat}{\stackrel 0 \rho}{}^{nim}{}_k)$. As a result, we find
\begin{eqnarray}
&&\nabla_i\nabla_n\left[F\left(\tau^{ni}{}_k + \tau^n{}_k{}^i - \tau^i{}_k{}^n\right)\right] =
F\tau^m{}_n{}^l R_{klm}{}^n,\label{nntau}\\
&&2\nabla_i\nabla_n\nabla_m(L_{\rm mat}{\stackrel 0 \rho}{}^{nim}{}_k) = R_{inlk}\nabla_m(L_{\rm mat}{\stackrel 0 \rho}{}^{inlm}).\label{nnn}
\end{eqnarray}
The Ricci identity was used during the derivation of these relations. Furthermore, making use of the Bianchi identity one can verify that
\begin{eqnarray}\label{nLrho}
2\nabla_i(L_{\rm mat}{\stackrel 0 \rho}{}_{lmn}{}^iR^{lmn}{}_k) &=&2\nabla_i(L_{\rm mat}{\stackrel 0 \rho}{}_{lmn}{}^i)R^{lmn}{}_k  \nonumber \\
&& + L_{\rm mat}{\stackrel 0 \rho}{}_{lmn}{}^i\nabla_kR^{lmn}{}_i.
\end{eqnarray}
Finally, (\ref{dPi}) is derived by combining (\ref{nntau}), (\ref{nnn}), (\ref{nLrho}) with (\ref{F1a}) and (\ref{q2a}). 

Equation (\ref{Pi}) generalizes our findings in \cite{Puetzfeld:Obukhov:2013} for the case of the matter with spin and a coupling function $F$ with an arbitrary dependence on the Riemann curvature. The conservation law (\ref{q2a}) can be recast into
\begin{equation}
{\nabla}{}_i\Sigma_k{}^i = - \tau^m{}_n{}^l R_{klm}{}^n - {\frac 1F}\left(\Sigma_k{}^i + L_{\rm mat}\delta^i_k\right)\nabla_iF.\label{consfin}
\end{equation}
Lowering the index in (\ref{q1a}) and antisymmetrizing, we find a modified conservation law of the total angular momentum
\begin{equation}\label{skewSgr}
F\Sigma_{[ij]} + {\nabla}{}_n\left(F\tau_{ij}{}^n\right) = 0.
\end{equation}
Thus, the divergence of the canonical energy-momentum tensor is balanced, in general, by the Mathisson-Papapetrou force plus the ``pressure-type'' nonminimal force proportional to the gradient of $F$. In the absence of spin, we recover the old result of \cite{Puetzfeld:Obukhov:2013}.

\section{Conclusion}\label{conclusion_sec}

In this paper, we have developed a unified framework for the discussion of Noether identities and conservation laws for a wide range of metric-affine gravitational theories. Our framework covers gauge theories of gravity (based on spacetime symmetry groups), and various so-called $f(R)$ models (and their generalizations), with and without minimal coupling of matter and gravity. 

The results obtained extend our earlier findings \cite{Puetzfeld:Obukhov:2013:2,Puetzfeld:Obukhov:2013:3}, which were derived for a more narrow class of models and geometries. The general conservation laws (\ref{cons1b}) and (\ref{cons2b}) form the starting point for the study of the equations of motion of extended test bodies. The latter should be necessarily built of matter with microstructure (such as the intrinsic hypermomentum, including spin, dilation and shear charges). 

\section*{Acknowledgements}
This work was supported by the Deutsche Forschungsgemeinschaft (DFG) through the grant LA-905/8-1/2 (D.P.). 

\bibliographystyle{unsrtnat}
\bibliography{consmag}

\begin{thebibliography}{26}
\providecommand{\natexlab}[1]{#1}
\providecommand{\url}[1]{\texttt{#1}}
\expandafter\ifx\csname urlstyle\endcsname\relax
  \providecommand{\doi}[1]{doi: #1}\else
  \providecommand{\doi}{doi: \begingroup \urlstyle{rm}\Url}\fi

\bibitem[{Einstein}(1956)]{Einstein:1956}
A.~{Einstein}.
\newblock \emph{{The meaning of relativity}}.
\newblock Princeton University Press, Princeton NJ, 5th revised edition, 1956.

\bibitem[{Schmidt}(2007)]{Schmidt:2007}
H.~J. {Schmidt}.
\newblock {Fourth order gravity: equations, history, and applications to
  cosmology}.
\newblock \emph{Int. J. Geom. Meth. Mod. Phys.}, 4:\penalty0 209, 2007.

\bibitem[{Straumann}(2008)]{Straumann:2008}
N.~{Straumann}.
\newblock {Problems with Modified Theories of Gravity, as Alternatives to Dark
  Energy}.
\newblock 2008.
\newblock URL \url{arXiv:0809.5148v1 [gr-qc]}.

\bibitem[{Nojiri} and {Odintsov}(2011)]{Nojiri:2011}
S.~{Nojiri} and S.~D. {Odintsov}.
\newblock {Unified cosmic history in modified gravity: from $F(R)$ theory to
  Lorentz non-invariant models}.
\newblock \emph{Phys. Rep.}, 505:\penalty0 59, 2011.

\bibitem[{Hehl} and {Kerlick}(1978)]{Hehl:1978}
F.~W. {Hehl} and G.~D. {Kerlick}.
\newblock {Metric-affine variational principles in general relativity. I.
  Riemannian spacetime}.
\newblock \emph{Gen. Relat. Grav.}, 9:\penalty0 691, 1978.

\bibitem[{Hehl} and {Kerlick}(1981)]{Hehl:1981}
F.~W. {Hehl} and G.~D. {Kerlick}.
\newblock {Metric-affine variational principles in general relativity. II.
  Relaxation of the Riemannian constraint}.
\newblock \emph{Gen. Relat. Grav.}, 13:\penalty0 1037, 1981.

\bibitem[{Sotiriou} and {Faraoni}(2010)]{Sotiriou:2010}
T.~P. {Sotiriou} and V.~{Faraoni}.
\newblock {$f(R)$ theories of gravity}.
\newblock \emph{Rev. Mod. Phys.}, 82:\penalty0 451, 2010.

\bibitem[{Blagojevi\'c}(2002)]{Blagojevic:2002}
M.~{Blagojevi\'c}.
\newblock \emph{{Gravitation and Gauge Symmetries}}.
\newblock IOP Publishing, London, 2002.

\bibitem[{Blagojevi\'c} and {Hehl}(2013)]{Hehl:2013}
M.~{Blagojevi\'c} and F.~W. {Hehl}.
\newblock \emph{{Gauge Theories of Gravitation. A Reader with Commentaries}}.
\newblock Imperial College Press, London, 2013.

\bibitem[{Schouten}(1954)]{Schouten:1954}
J.~A. {Schouten}.
\newblock \emph{{Ricci-Calculus. An introduction to tensor analysis and its
  geometric applications}}.
\newblock Springer, Berlin, 2nd edition, 1954.

\bibitem[{Einstein}(1921)]{Einstein:1921}
A.~{Einstein}.
\newblock {Geometrie und Erfahrung}.
\newblock \emph{Sitzungsber. preuss. Akad. Wiss.}, 1:\penalty0 123, 1921.

\bibitem[{Hehl} et~al.(1995){Hehl}, {McCrea}, {Mielke}, and
  {Ne'eman}]{Hehl:1995}
F.~W. {Hehl}, J.~D. {McCrea}, E.~W. {Mielke}, and Y.~{Ne'eman}.
\newblock {Metric-affine gauge theory of gravity: Field equations, Noether
  identities, world spinors, and breaking of dilation invariance}.
\newblock \emph{Phys. Rep.}, 258:\penalty0 1, 1995.

\bibitem[{Bertolami} et~al.(2007){Bertolami}, {B\"ohmer}, {Harko}, and
  {Lobo}]{Bertolami:etal:2007}
O.~{Bertolami}, C.~G. {B\"ohmer}, T.~{Harko}, and F.~S.~N. {Lobo}.
\newblock {Extra force in $f(R)$ modified theories of gravity}.
\newblock \emph{Phys. Rev. D.}, 75:\penalty0 104016, 2007.

\bibitem[{Puetzfeld} and {Obukhov}(2013{\natexlab{a}})]{Puetzfeld:Obukhov:2013}
D.~{Puetzfeld} and Yu.~N. {Obukhov}.
\newblock {Covariant equations of motion for test bodies in gravitational
  theories with general nonminimal coupling}.
\newblock \emph{Phys. Rev. D}, 87:\penalty0 044045, 2013{\natexlab{a}}.

\bibitem[{Sciama}(1962)]{Sciama:1962}
D.~W. {Sciama}.
\newblock {On the analogy between charge and spin in general relativity}.
\newblock \emph{``Recent Developments in General Relativity'', Festschrift for
  L. Infeld (Pergamon Press, Oxford; PWN, Warsaw)}, page 415, 1962.

\bibitem[{Kibble}(1961)]{Kibble:1961}
T.~W.~B. {Kibble}.
\newblock {Lorentz invariance and the gravitational field}.
\newblock \emph{J. Math. Phys.}, 2:\penalty0 212, 1961.

\bibitem[{Hehl} et~al.(1976{\natexlab{a}}){Hehl}, {Kerlick}, and {von der
  Heyde}]{Hehl:1976a}
F.~W. {Hehl}, G.~D. {Kerlick}, and P.~{von der Heyde}.
\newblock {On hypermomentum in general relativity. I. The notion of
  hypermomentum}.
\newblock \emph{Zeits. Naturforsch.}, 31a:\penalty0 111, 1976{\natexlab{a}}.

\bibitem[{Hehl} et~al.(1976{\natexlab{b}}){Hehl}, {Kerlick}, and {von der
  Heyde}]{Hehl:1976b}
F.~W. {Hehl}, G.~D. {Kerlick}, and P.~{von der Heyde}.
\newblock {On hypermomentum in general relativity. II. The geometry of
  spacetime}.
\newblock \emph{Zeits. Naturforsch.}, 31a:\penalty0 524, 1976{\natexlab{b}}.

\bibitem[{Hehl} et~al.(1976{\natexlab{c}}){Hehl}, {Kerlick}, and {von der
  Heyde}]{Hehl:1976c}
F.~W. {Hehl}, G.~D. {Kerlick}, and P.~{von der Heyde}.
\newblock {On hypermomentum in general relativity. III. Coupling hypermomentum
  to geometry}.
\newblock \emph{Zeits. Naturforsch.}, 31a:\penalty0 823, 1976{\natexlab{c}}.

\bibitem[{Ponomariev} and {Obukhov}(1982)]{Obukhov:1982}
V.~N. {Ponomariev} and Yu.~N. {Obukhov}.
\newblock {The generalized Einstein-Maxwell theory of gravitation}.
\newblock \emph{Gen. Relat. Grav.}, 14:\penalty0 309, 1982.

\bibitem[{Vassiliev}(2005)]{Vassiliev:2005}
D.~{Vassiliev}.
\newblock {Quadratic metric-affine gravity}.
\newblock \emph{Ann. Phys. (Leipzig)}, 14:\penalty0 231, 2005.

\bibitem[{Sotiriou} and {Liberati}(2007)]{Sotiriou:2007}
T.~P. {Sotiriou} and S.~{Liberati}.
\newblock {Metric-affine $f(R)$ theories of gravity}.
\newblock \emph{Ann. Phys. (N.Y.)}, 322:\penalty0 935, 2007.

\bibitem[{Vitagliano} et~al.(2011){Vitagliano}, {Sotiriou}, and
  {Liberati}]{Vitagliano:2011}
V.~{Vitagliano}, T.~P. {Sotiriou}, and S.~{Liberati}.
\newblock {The dynamics of metric-affine gravity}.
\newblock \emph{Ann. Phys. (N.Y.)}, 326:\penalty0 1259, 2011.

\bibitem[{Obukhov} and {Rubilar}(2007)]{Obukhov:2007}
Yu.~N. {Obukhov} and G.~F. {Rubilar}.
\newblock {Invariant conserved currents in gravity theories: Diffeomorphisms
  and local gauge symmetries}.
\newblock \emph{Phys. Rev. D}, 76:\penalty0 124030, 2007.

\bibitem[{Obukhov} and {Puetzfeld}(2013)]{Puetzfeld:Obukhov:2013:2}
Yu.~N. {Obukhov} and D.~{Puetzfeld}.
\newblock {Conservation laws in gravitational theories with general nonminimal
  coupling}.
\newblock \emph{Phys. Rev. D}, 87:\penalty0 081502(R), 2013.

\bibitem[{Puetzfeld} and
  {Obukhov}(2013{\natexlab{b}})]{Puetzfeld:Obukhov:2013:3}
D.~{Puetzfeld} and Yu.~N. {Obukhov}.
\newblock {Equations of motion in gravity theories with nonminimal coupling: A
  loophole to detect torsion macroscopically?}
\newblock \emph{Phys. Rev. D}, 88:\penalty0 064025, 2013{\natexlab{b}}.

\end{thebibliography}

\end{document}